\documentclass[%
 preprint,
 amsmath,amssymb,
 aps, physrev,
]{revtex4-2}

\usepackage{graphicx}
\usepackage{dcolumn}
\usepackage{bm}
\usepackage{xcolor} 
\usepackage[normalem]{ulem} 

\begin{document}

\preprint{APS/123-QED}

 
\title{Breakdown of Kolmogorov Scaling and Modified Energy Transfer in Bubble-Laden Turbulence}

\author{Andrea Montessori}

\affiliation{%
Department of Civil, Computer Science and Aeronautical Technologies Engineering,  Roma Tre University, Rome, Italy 
}%
 \email{Contact author: andrea.montessori@uniroma3.it}

\author{Marco Lauricella}

\affiliation{%
  Istituto per le Applicazioni del Calcolo, Consiglio Nazionale delle Ricerche,Via dei Taurini 19,Rome,00185, Italy
}%

\author{Aritra Mukherjee}

\affiliation{%
Department of Energy and Process Engineering, Norwegian University of Science and Technology, Trondheim, Norway 
}%

\author{Luca Brandt}

\affiliation{%
Dipartimento di Ingegneria dell’Ambiente, del Territorio e delle Infrastrutture, Politecnico di Torino, Torino, Italia 
}%

\date{\today}

\begin{abstract}

We investigate the effect of a dispersed bubble phase on forced homogeneous and isotropic turbulence using high-resolution high-performance simulations based on the lattice Boltzmann method. While the classical Kolmogorov energy cascade is largely preserved when considering the system as a whole, a phase-specific analysis reveals striking deviations from the classical turbulence scaling. In particular, the gas phase exhibits significant departures from Kolmogorov’s predictions, whereas the continuous liquid phase retains a turbulence structure consistent with classical expectations  \textcolor{black}{up to $24\%$ in gas volume fractions}. These findings suggest that, despite the presence of a dispersed phase, the global energy transfer remains close to a universal behavior. At the same time, phase-specific interactions are shown to introduce modifications to the turbulent dynamics at small scales. In particular, the gas-phase exhibits a nearly flat spectrum at low wave numbers followed by a $k^{-3}$ scaling at intermediate scales pointing to the presence of patterns of localized bursts  uniformly distributed
between two finite wavelengths. 
Our results aim at deepening the understanding of multiphase turbulence, particularly in the context of energy transfer mechanisms and phase interactions in bubble-laden flows. This study provides a framework for future investigations into the fundamental properties of multiphase turbulence and its implications for environmental, atmospheric, and industrial flows.

\end{abstract}

\maketitle


\section{Introduction}

Multiphase turbulent flows are ubiquitous in both natural and industrial settings, playing a crucial role in a broad range of applications, from oceanic and atmospheric processes to chemical reactors and energy systems \cite{veron2015ocean,chandran2015study,angulo2020influence}. Among these, bubble-laden turbulence is particularly relevant, as it governs key phenomena such as wave breaking \cite{christensen2001large,iafrati2009numerical}, gas transfer at the ocean–atmosphere interface \cite{veron2015ocean,bigg2008composition}, cavitation \cite{caupin2006cavitation}, and industrial mixing processes \cite{bureiko2015current}. Despite its widespread importance, the fundamental mechanisms underpinning turbulence modulation by dispersed bubbles remain only partially understood. The presence of an additional phase introduces complex interfacial interactions \cite{montessori2019mesoscale,tiribocchiPhysRep}, modifies energy transfer across scales \cite{crialesi2022modulation}, and alters small-scale dynamics in ways that are challenging to predict using classical turbulence theories.

One of the most prominent challenges in studying bubble-laden turbulence stems from its intrinsic multiscale and multiphysics nature. The interaction between the continuous phase and the dispersed gas phase leads to subtle but fundamental deviations from classical Kolmogorov turbulence \cite{Pope_2000}, requiring novel approaches to accurately characterize energy transfer mechanisms. Understanding these effects is essential for improving predictive models in climate science, where bubble-induced turbulence affects the dispersion of aerosols and microplastics \cite{allen2020examination}, as well as in industrial applications, such as the optimization of bubble column reactors and cavitation mitigation strategies \cite{kantarci2005bubble,plesset1977bubble}, to cite some prominent examples.

Due to the inherent complexity of these flows, direct experimental measurements are often limited, particularly at the small scales where interphase interactions dominate \cite{annurevlohsebubble2020}. As a result, numerical simulations play a fundamental role in uncovering the underlying physics. However, accurately resolving turbulence in the presence of bubbles requires high-resolution computations, advanced numerical schemes, and high-performance computing (HPC) frameworks capable of handling large density and viscosity contrasts, as well as the complex interactions between phases in turbulent environments.

In this work, a recently developed HPC implementation \cite{montessori2025thread,montessori2024order,lauricella2025acclb} of the lattice Boltzmann (LB) method \cite{succi,montessori2018lattice,kruger2017lattice} is employed to investigate the effect of dispersed bubbles on homogeneous isotropic turbulence (HIT). By performing numerical experiments across different bubble volume fractions, density ratios, and viscosity ratios, we provide a detailed analysis of energy transfer mechanisms at both global and phase-specific levels. Our results reveal that, while the overall energy spectrum maintains a Kolmogorov-like behavior, the gas and liquid phases exhibit distinct turbulence characteristics, with the gas phase deviating significantly from classical scaling laws. In particular, we observe that bubble-induced momentum transfer redistributes turbulent energy, leading to the emergence of a non-classical cascade within the dispersed phase.

These findings offer new insights into the role of gas–liquid interactions in turbulent energy transfer and highlight the necessity of phase-specific analyses in multiphase turbulence studies, which are also fundamental to improving two-fluid models. The results presented here confirm that a robust computational framework is crucial for future investigations into bubble-laden turbulence and open the way for improved models in environmental and industrial multiphase flow applications.

\section{Methods}

\subsection{Navier-Stokes equations for multiphase flows with interfaces}

The multiphase system considered in this study is modeled by solving the continuity and momentum equations using the LB method. These equations are given by:
\begin{equation}
\frac{\partial \rho}{\partial t} + \nabla \cdot (\rho \mathbf{u}) = 0
\end{equation}
\begin{equation} \label{NSeq}
\frac{\partial (\rho \mathbf{u})}{\partial t} + \nabla \cdot (\rho \mathbf{u}  \mathbf{u}) = -\nabla p + \nabla \cdot \left( \rho \nu \left[ \nabla \mathbf{u} + (\nabla \mathbf{u})^\top \right] \right) + \mathbf{F}^s
\end{equation}
where \( \mathbf{u} \) represents the velocity field, \( \rho \) the density, \( p \) the macroscopic pressure, \( \nu \) the kinematic viscosity, and \( \mathbf{F^s} \) the surface tension force. Greek indices indicate Cartesian components of vectors and tensors. 

The surface tension force, \( \mathbf{F^s} \), can generally be written as the divergence of a capillary stress tensor:
\begin{equation}
\mathbf{F}^s = \nabla \cdot \left( \sigma (\mathbf{I} - \mathbf{n} \otimes \mathbf{n}) \right)
\end{equation}
where \( \mathbf{n} \) is the local normal to the interface and \( \sigma \) is the surface tension coefficient, here assumed constant (zero tangential stresses). 

\subsubsection{Allen-Cahn Equation for Interface Tracking}

The interface dynamics is captured through the evolution of a phase field, \( \phi \), governed by the conservative Allen-Cahn equation:
\begin{equation}
\frac{\partial \phi}{\partial t} + \mathbf{u} \cdot \nabla \phi = D \nabla^2 \phi - \kappa \nabla \cdot \left[ \phi(1 - \phi) \mathbf{n} \right]
\end{equation}
where \( D \) is the interface diffusivity, and \( \kappa = 4 D/\delta \), with \( \delta \) denoting the interface width. In this model, \( \phi \) takes values within the range \( \{ 0,1 \} \), such that the interface between the two immiscible fluids is located at \( \phi_0=0.5 \). It has been demonstrated \cite{kim2005continuous} that the equilibrium profile for an interface positioned at \( x_0 \) follows:
\begin{equation}
    \phi(x, y, z)=\phi_0 \pm \frac{\phi_H - \phi_L}{2}\tanh\Bigg( \frac{|x-x_0|+|y-y_0|+|z-z_0|}{\delta}\Bigg).
\end{equation}
The interface tracking equation is dynamically coupled to the Navier-Stokes equations through the advection term. 

To enhance the accuracy of the model in the presence of strong advection, the advection term in Eq. (4) is discretized using a fifth-order Weighted Essentially Non-Oscillatory (WENO-5) scheme\cite{shu2003high}. The WENO-5 scheme provides high-order accuracy while effectively handling sharp interface transitions and minimizing numerical dissipation. This approach ensures stable and accurate advection of the phase field variable $\phi$, particularly in turbulent multiphase flows where sharp interfaces need to be captured with minimal spurious oscillations. See \cite{montessori2024order, lauricella2025thread} for validations of the present two-phase flow solver at high density ratio. 

\subsection{Thread-safe lattice Boltzmann model}

The Navier-Stokes equations for single-phase flows can be efficiently solved using the LB method. This section provides a brief overview of a thread-safe LB framework, which integrates an efficient LB strategy with a regularized LB approach \cite{montessori2023thread,lauricella2025acclb,montessori2025thread}. 

The Boltzmann equation, discretized over a velocity space of $q$ discrete vectors, is given by:
\begin{equation} \label{lbe}
    f_i(\mathbf{x} + \mathbf{c}_i \Delta t, t + \Delta t) = f_i(\mathbf{x}, t) + \omega (f_i^{\text{eq}} - f_i) + (1-\frac{\omega}{2}) S_i
\end{equation}
where $f_i$ are the distribution functions, $f_i^{eq}$ their equilibrium counterpart, and $S_i$ represents an external forcing term. The relaxation rate is controlled by $\omega$, which is linked to the kinematic viscosity of the fluid via the relation $\nu=c_s^2(1/\omega - 0.5)$ ( see \cite{succi,kruger2017lattice,montessori2018lattice} for details on the LB method).
By decomposing $f_i=f_i^{eq} + f_i^{neq}$, the post-collision distribution can be rewritten as:
\begin{equation}\label{pushLB}
    f_i(\mathbf{x} + \mathbf{c}_i \Delta t, t + \Delta t) = f_i^{\text{eq}} + (1 - \omega) f_i^{\text{neq}} + (1-\frac{\omega}{2})S_i
\end{equation}

In the thread-safe approach, both the equilibrium and non-equilibrium parts are reconstructed from the macroscopic hydrodynamic fields, eliminating the need to read the full set of lattice populations. This prevents race conditions arising from non-local memory access \cite{montessori2023thread,montessori2024order}. The equilibrium distributions take the form:
\begin{align}\label{eqts}
    f_i^{\text{eq}} = w_i \left( p^* + \frac{\mathbf{c}_i \cdot \mathbf{u}}{c_s^2} + \frac{(\mathbf{c}_i \mathbf{c}_i - c_s^2 \mathbf{I}) : \mathbf{u} \mathbf{u}}{2 c_s^4} \right)
\end{align}
where $p^*=p/(\rho c_s^2)$ is a dimensionless pressure. The non-equilibrium part is reconstructed up to second order as:
\begin{equation} \label{noneqts}
    f_i^{\text{neq}} = \frac{\mathbf{c}_i \mathbf{c}_i : \mathbf{a}_{\text{neq}}}{2 c_s^4}
\end{equation}
$\mathbf{a}_{\text{neq}}$ the non-equilibrium stress tensor.

After the collision step, the macroscopic pressure and velocity fields are retrieved using:
\begin{equation} \label{moments0}
    p^* (\mathbf{x},t)=\sum_i f_i(\mathbf{x},t),
\end{equation}
\begin{equation} \label{moments1}
    u(\mathbf{x},t)=\sum_i f_i(\mathbf{x},t)\mathbf{c}_{i} + \frac{1}{2}\sum_i S_i.
\end{equation}

The forcing term $S_i$ follows the Guo forcing scheme \cite{guo2002discrete}:
\begin{equation} \label{guoforce}
    S_i = w_i \left( \frac{\mathbf{c}_i - \mathbf{u}}{c_s^2} + \frac{\mathbf{c}_i \cdot \mathbf{u}}{c_s^4} \, \mathbf{c}_i \right) \cdot \mathbf{F}
\end{equation}
where $\mathbf{F}$ represents the external force field. 

\subsection{Extension for high-density and viscosity contrasts}

Additional forces are introduced to capture surface tension effects and inhomogeneous density contributions to model multiphase flows. The total force is given by:
\begin{equation}
    \mathbf{F}=\mathbf{F}^s + \mathbf{F}^p +\mathbf{F}^\nu,
\end{equation}
where the surface tension force is computed as follows \cite{kim2005continuous}:
\begin{equation}
\mathbf{F}^s=-\gamma\kappa\nabla\phi|\nabla\phi|
\end{equation}
.

The pressure force is \cite{fakhari2017improved}:
\begin{equation}\label{press_corr}
\mathbf{F}^p = -p^* c_s^2 \nabla \rho
\end{equation}
where $\rho=\rho_l \phi+(1-\phi)\rho_g$. Since the local pressure is $p=p^*\rho c_s^2$, its gradient includes both embedded and additional terms, requiring the explicit inclusion of $\mathbf{F}^p$.

The viscous force is given by \cite{fakhari2017improved}:
\begin{equation}\label{visc_corr}
\mathbf{F}^\nu = -\frac{\nu \omega}{c_s^2 \Delta t} \left[ \sum_i (f_i - f_i^{\text{eq}}) \, \mathbf{c}_i  \mathbf{c}_i \right] \nabla \rho
\end{equation}


This framework enables stable and efficient simulation of multiphase flows with high-density and viscosity contrasts while maintaining thread safety through regularization \cite{montessori2024order}.

\section{Results}

\subsection{Single-phase homogeneous isotropic turbulence: spectra and single point statistics}

We first proceed to analyze single-phase HIT energy and pressure spectra and selected single-point statistics so to assess the accuracy of \textit{accLB} \cite{montessori2024order, lauricella2025acclb} to perform fully-developed turbulence simulations. 

For both single-phase and bubble-laden simulations the homogeneous isotropic turbulent state is maintained through large-scale energy injection via the ABC (Arnold–Beltrami–Childress) forcing. This deterministic, divergence-free forcing scheme introduces a time-independent velocity field that acts as a steady source of energy at the largest scales. Specifically, the forcing takes the form:
\[
\mathbf{f}(\mathbf{x}) = A
\begin{pmatrix}
\sin(k_f z) + \cos(k_f y) \\
\sin(k_f x) + \cos(k_f z) \\
\sin(k_f y) + \cos(k_f x)
\end{pmatrix},
\]
where \(A\) is the forcing amplitude and \(k_f\) the forcing wavenumber. In this work, the forcing amplitude and the forcing wavenumber have been set to $2\cdot10^{-7}$(expressed in units of lattice) and $2$, respectively. The ABC forcing guarantees a sustained turbulent regime by balancing viscous dissipation with energy input, without introducing significant anisotropy, thereby providing an efficient and widely adopted method to achieve statistically stationary HIT in periodic domains \cite{singh2024comparison}.

To note here that, the forcing is applied only in the bulk liquid phase, as in \cite{Rivière_Mostert_Perrard_Deike_2021}.

Two cases with increasing Taylor-scale Reynolds number ($Re_\lambda=u_{rms}\lambda/\nu$, $\lambda=\sqrt{\frac{15\nu}{\varepsilon}}u_{rms}$), \(Re_\lambda\)= 150 and 230, have been considered. The simulations were conducted on a \(512^3\) Cartesian uniform grid.
The \(Re_\lambda = 150\) case qualifies as a Direct Numerical Simulation (DNS), with the Kolmogorov length scale approximately \(\Delta x\), ensuring that the smallest turbulent scales are resolved. On the other hand, the \(Re_\lambda = 230\) case operates slightly outside the strict DNS regime, with a Kolmogorov length scale of \(0.4\Delta x\). 

All the simulations have been run on Leonardo supercomputer \cite{turisini2024leonardo} at CINECA, Italy. The details of the implementation of \textit{accLB} and the related performances on both single and multiGPU can be found in \cite{lauricella2025acclb}. 

\begin{figure}
    \centering
    \includegraphics[width=1.\linewidth]{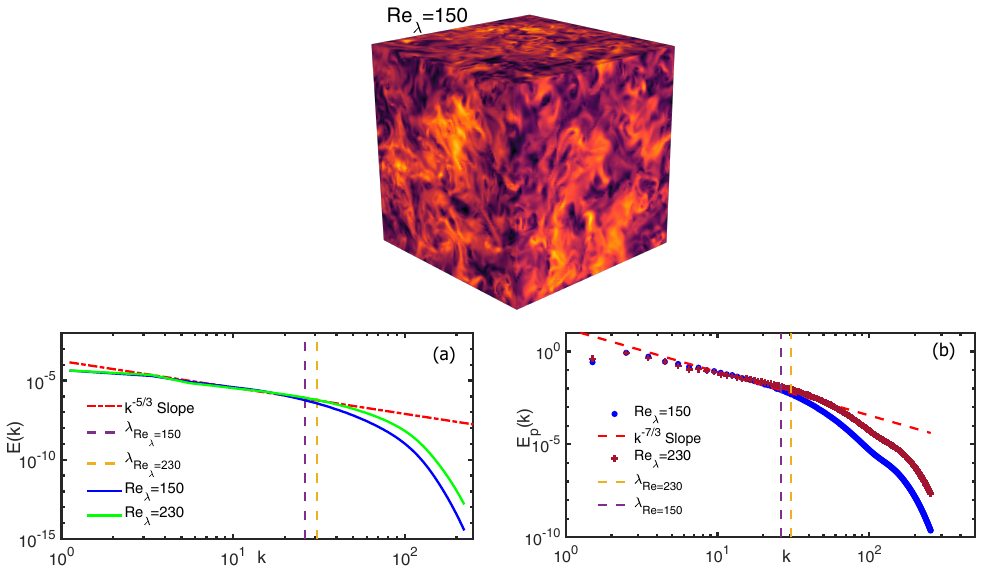}
    \caption{Volume representation of the instantaneous velocity field in a homogeneous single-phase turbulent flow at $Re_\lambda=150$. The domain size is $512^3$. In panel (a): the energy spectrum for two different Taylor Reynolds lambda, namely $150$ and $230$. The expected $-5/3$ law in the inertial range is well reproduced across a wide range of wavenumbers. In panel (b), the pressure spectra for the same Taylor Reynolds number are shown to follow the $-7/3$ scaling law within the inertial range. }
    \label{fig:spectraSC}
\end{figure}

Figure~\ref{fig:spectraSC}(a) compares the energy spectra for both cases with the classical \(-5/3\) slope expected in the inertial sub-range \cite{Pope_2000, frisch1995turbulence}. The agreement is evident and this stands as a first fundamental proof of the ability of \textit{accLB} to capture energy transfer across scales in single-phase turbulent flows accurately.

Moreover, the pressure fluctuations spectra are displayed and compared against the expected \(-7/3\) power-law \cite{zhao2016} in figure~\ref{fig:spectraSC}(b). For both the pressure and energy spectra, the simulation at the highest Reynolds number exhibits the theoretical scaling over a broader range than the $Re_\lambda=150$ case, reflecting the extended inertial range and increased scale separation. 
As expected, the spectra start to deviate from the $-5/3$ and $-7/3$ power-laws at the larger scales, in proximity of the Taylor length scale, which is denoted by the vertical dashed lines in the figure.

\begin{figure}
    \centering
    \includegraphics[width=0.8\linewidth]{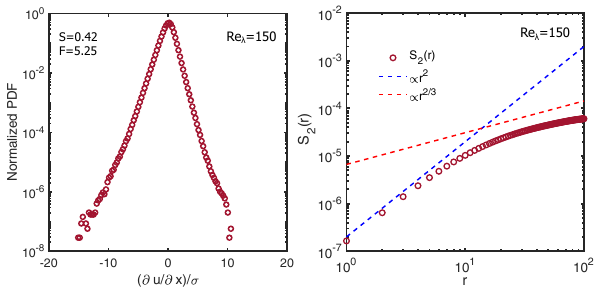}
    \caption{Panel(a) reports the normalized probability density function (PDF) of the longitudinal velocity gradients at $Re_\lambda=150$, with the values of skewness and kurtosis in the legend. Panel(b): second-order structure function compared with two theoretical scaling laws, namely the $r^2$ scaling in the dissipation range and the Kolmogorov scaling, $r^{2/3}$, in the inertial range.}
    \label{fig:spsSC}
\end{figure}

Single-point statistics of the two HIT flows in scrutiny are shown in Figure \ref{fig:spsSC}. In particular, panel(a) reports the probability density function (PDF) of the longitudinal velocity gradients for the flow at $Re_\lambda=150$.  
As expected, the PDF is far from Gaussian and slightly skewed, with a skewness factor, $S=0.42$ ($S=-<(\partial u/\partial x)^3>/<(\partial u/\partial x)^2>^{3/2}$), and a flatness factor $F=5.25$ ($F=<(\partial u/\partial x)^4>/<(\partial u/\partial x)^2>^{2}$), closely aligning with previous findings (see \cite{ishihara2009} as a reference)  at similar values of the Taylor Reynolds number. The first index is a typical hallmark of the emergence of anisotropic effects at small scales, even in globally isotropic turbulence, driven by the presence of persistent vortex stretching. In contrast, the non-Gaussian  value of the  flatness (larger than $3$) indicates the presence of  intermittency, i.e., rare but intense fluctuations of the velocity gradients \cite{mccomb1995theory}.

In panel (b) the second-order structure function, $S_2(r)=<[u(x_i+r_i)-u(x_i)]^2>$, is reported and compared with two theoretical scaling laws \cite{mccomb1995theory}: the $r^2$ scaling in the dissipation range (small values of r, fine-scale pseudo-laminar regime) and the Kolmogorov scaling, $r^{2/3}$, in the inertial range (larger values of $r$ corresponding to larger-scale structures). 
The same agreement with the existing literature is seen for the results of the simulation at higher Taylor Reynolds number case (i.e., $Re_\lambda=230$). For the sake of conciseness, results have only been shown for the case $Re_\lambda=150$.  

\subsection{Bubble-induced turbulence modulation in homogeneous isotropic turbulence}

In the following, we examine the modulation of HIT due to the presence of a dispersed gas phase. To this aim, we performed simulations across a range of bubble volume fractions (\(\phi_B\)), viscosity ratios, and density ratios. In particular, the focus is on the impact of gas-liquid interfaces on higher-order turbulence statistics and their connection to the modified turbulent energy spectra.

\begin{figure}
\centering
\includegraphics[width=0.9\linewidth]{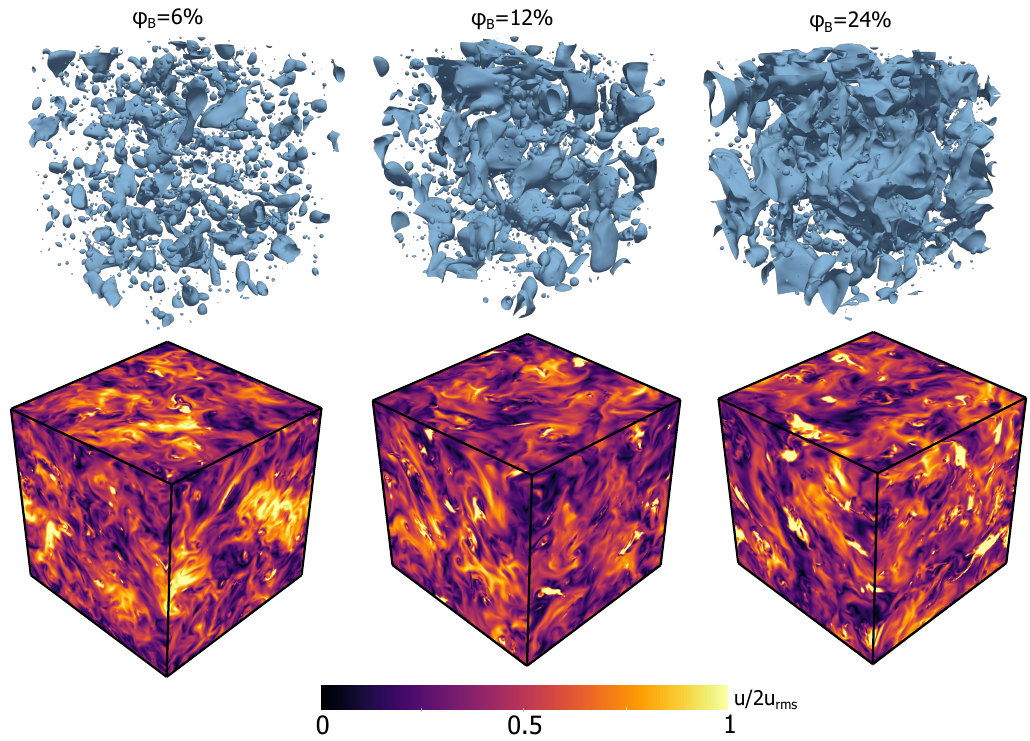}
\caption{Instantaneous snapshots of the dispersed bubble phase and velocity field for three different bubble volume fractions, $\phi_B$. The snapshots are taken after HIT has reached an averaged steady state. The density and viscosity ratios are set to $100$ and $1$, respectively.}
\label{fig:voids}
\end{figure}

The simulations were performed on a \(512^3\) domain, consistent with the single-phase reference case. The baseline physical parameters consist of a viscosity ratio  \(\nu_G / \nu_L = 1\) and a density ratio of \(\rho_L / \rho_G = 100\). The void fraction was varied from \(0\%\) (single-phase flow) to \(24\%\), as shown in Figure \ref{fig:voids}, reporting instantaneous snapshots of the dispersed bubbly phase and of the velocity field for the volume fractions investigated. Additionally, as for the single-phase flow, we investigate two different Taylor-scale Reynolds numbers, \(Re_\lambda = 150\) and \(230\). The surface tension has been set to $\sigma=5\cdot10^{-3}$(expressed in lattice units) delivering a turbulent Weber number $We=\rho_L u_{rms}^2\lambda/\sigma\sim3$. 

For the present set of physical parameters, the Hinze length scale has been inferred \textit{a posteriori} by employing the formula proposed by Deane and Stokes \cite{deane2002scale}, $l_H = 2^{-8/5} \, \varepsilon^{-2/5} \left( \frac{\gamma \, We_c}{\rho} \right)^{3/5}$ assuming a critical Weber number of \( We_c \sim 4 \). This choice is in agreement with the findings of Mart\'{\i}nez-Baz\'an \textit{et al.}~\cite{martinez1999breakup}, who reported similar critical conditions for bubble fragmentation in fully developed turbulence. The resulting estimate is \( l_H \approx 24 \), expressed in lattice units. 

\subsubsection{Effect of bubble volume fractions}

\begin{figure}
\centering
\includegraphics[scale=0.7]{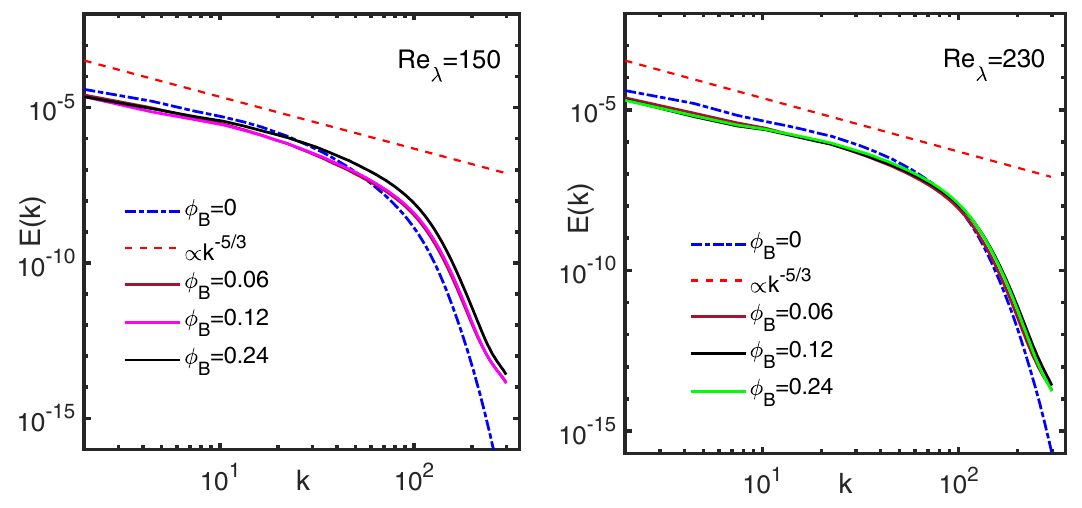}
\caption{Energy spectra at $Re_\lambda=150$ and $230$ for different bubble volume fractions, $\phi_B=0.06$, $0.12$, and $0.24$.}
\label{fig:E_k_comp_phi}
\end{figure}

The analysis of energy spectra from HIT simulations, with and without bubbles, reveals that the presence of bubbles leads, in the first place, to a redistribution of energy across scales. As shown in Figure~\ref{fig:E_k_comp_phi}, the dispersed phase causes a certain decrease in the energy content at low wavenumbers ($k$), which is almost independent of the gas volume fraction (i.e., $\phi_B$), and an increase in the energy content at the highest $k$, corresponding to small-scale structures, which is more pronounced when increasing $\phi_B$.

As observed in earlier works for emulsions with unitary density ratio \cite{crialesi2023intermittency, crialesi2022modulation}, such redistribution is to be attributed to the role of the surface tension force, which removes energy at the largest scale by deforming and breaking the larger bubbles and introduces additional energy at small scales during bubble merging and interface relaxation. As demonstrated below, the energy increase at high $k$ points to the presence of positive time-averaged work of the surface tension force at high wavenumbers, which is effectively injecting energy into small-scale turbulence. In \cite{crialesi2022modulation,pandey2022turbulence,pandey2020liquid}, it is shown that surface tension effects enhance energy content at small scales by modifying the energy transfer pathways of the two-phase, emulsion system.

Note that, as pointed out above, the spectral modifications weakly depend on the bubble volume fraction for the parameters of our simulations, except for the smallest active scales. As demonstrated by the spectra in Figure~\ref{fig:E_k_comp_phi}, simulations with increasing bubble volume fractions ($\phi_B = 0.06$, $0.12$, and $0.24$) at Taylor Reynolds numbers of $Re_\lambda = 150$ and $230$ exhibit similar trends of the energy cascade. This consistency suggests that, at least for the range of volume fractions under scrutiny, the presence of bubbles, rather than their concentration, plays the dominant role in altering the energy transfer dynamics and that the mass fraction of the dispersed phase, rather than the volume fraction, plays an important role for the turbulence modulation. Indeed, in the case of emulsions (i.e. unitary density ratio)  in \cite{crialesi2022modulation} the increase of the turbulence modulation with the volume fraction of the dispersed phase is more pronounced, especially at the larger scales.

\begin{figure}
    \centering
    \includegraphics[width=0.5\linewidth]{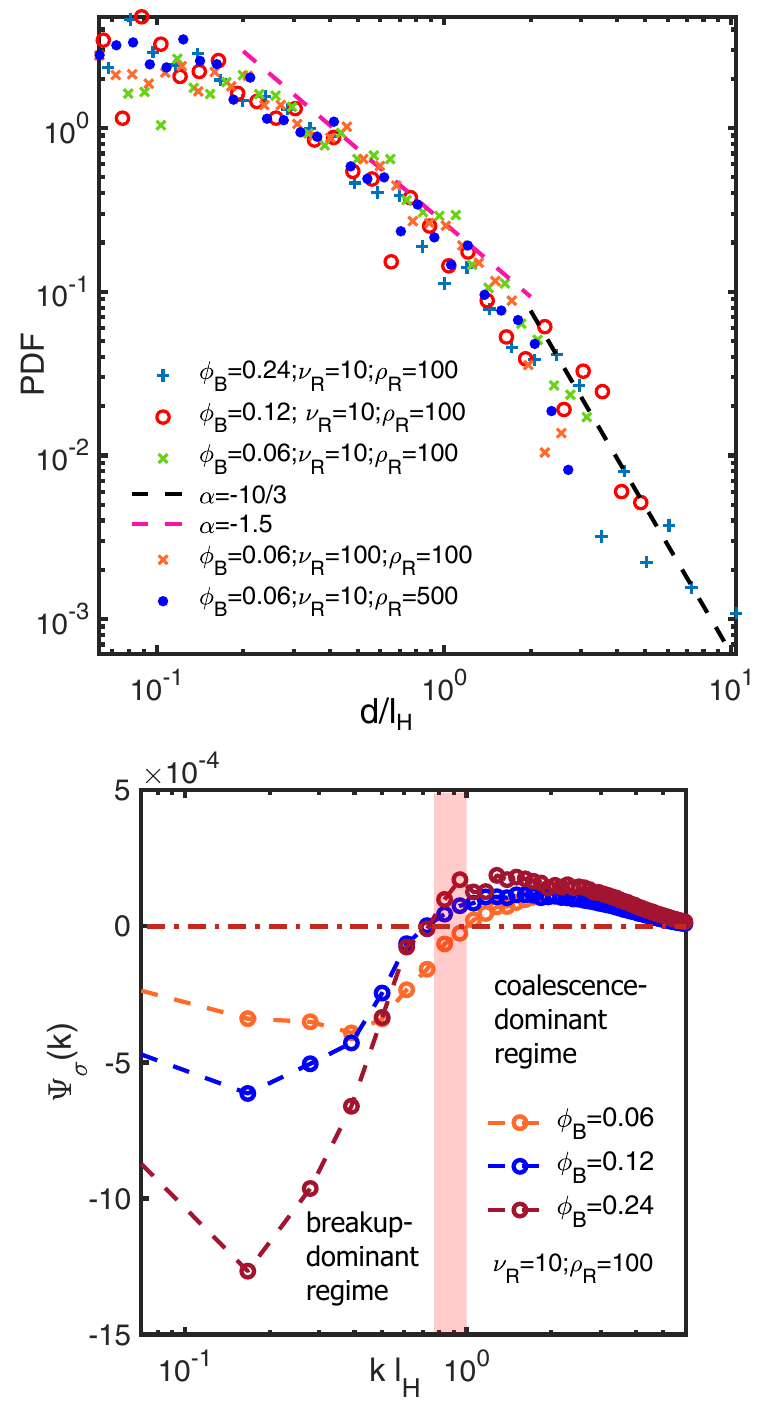}
    \caption{(a) Probability density functions of equivalent bubble diameters \(d\), normalized by the Hinze scale \(l_H\), for various flow parameters. (b) Net surface tension contribution \(\Psi(k)\) as a function of wavenumber \(k\), normalized by \(l_H\). The shadowed area marks the transition region \(k \sim 1/l_H\) separating the breakup- and coalescence-dominated regimes. $\nu_R$ and $\rho_R$ denotes the viscosity and density ratios.}
    \label{fig:pdf_diam}
\end{figure}

Further insights on the energy redistribution across scales can be obtained by examining the statistical distribution of bubble sizes in our simulations. We define the bubble size as the diameter of a sphere of the same volume of the bubble under consideration. The probability density functions (PDFs) of the equivalent bubble diameters, normalized by the Hinze scale \( d/l_H \), are reported in Figure~\ref{fig:pdf_diam}(a) for various values of the bubble volume fraction \(\phi_B\), density ratio \(R = \rho_l/\rho_g\), and viscosity ratio.
Despite variations in these control parameters, all cases exhibit a two-regime scaling: a \(-3/2\) power law for \(d/l_H \lesssim 1\), and a steeper \(-10/3\) decay for \(d/l_H \gtrsim 1\). The latter behavior reflects a universal fragmentation mechanism driven by turbulent stresses, which dominate over surface tension beyond the Hinze scale. The transition near \(d \sim l_H\) signals the scale at which turbulence becomes strong enough to overcome capillary resistance and break up bubbles, in agreement with previous studies~\cite{crialesi2022modulation,deane2002scale,dodd2016interaction}. Notably, the persistence of this scaling across different \(\phi_B\) and \(R\) values highlights the self-similar character of bubble size distributions in statistically steady bubble-laden turbulence.
The presence of two distinct regimes for bubble dynamics-- coalescence-dominated for \(d < l_H\) and breakup-dominated for \(d > l_H\), is further corroborated by the spectral distribution of the interfacial energy transfer shown in Fig.~\ref{fig:pdf_diam}(b).
 The figure shows the scale-dependent surface tension work \(\Psi_\sigma(k)=<\mathbf{u} \cdot \mathbf{F}^s>\) as a function of the wavenumber \(k\) for the baseline case at different bubble volume fractions. In line with the analysis reported in \cite{crialesi2023interaction,saeedipour2025enstrophy}, we observe that \(\Psi_\sigma(k)\) changes sign near \(k \sim 1/l_H\):
 surface tension drains energy at larger scales, (\(k < 1/l_H\)), associated with bubble deformation and breakup while it
  injects energy into the carrier flow at small scales (\(k > 1/l_H\)), indicative of coalescence and interface relaxation.
  As shown in \cite{crialesi2023interaction,saeedipour2025enstrophy}, this crossing point defines the Hinze scale dynamically and matches the scale identified in the PDFs of the droplets size. Indeed, the sign inversion in $\Psi_\sigma(k)$ close to the Hinze scale implies a physical mechanism whereby surface tension competes with inertial forces: breaking large bubbles while smoothing small interfacial disturbances. This dual role reflects how surface tension extracts energy from large-scale bubble deformation and releases it at smaller scales through rapid interface relaxation events, such as coalescence-driven recoil, which contribute to localized bursts of kinetic energy and enhanced intermittency in the gas phase.
The alignment between the energy-sign inversion in $\Psi_\sigma(k)$ and the slope transition in \(P(d)\) supports the interpretation of the Hinze scale as the boundary between non-local coalescence-driven dynamics and local fragmentation cascades with the energy increase at high $k$ observed in figure \ref{fig:E_k_comp_phi}.

\subsubsection{Budget of turbulent kinetic energy}
\begin{figure}
    \centering
    \includegraphics[width=0.55\linewidth]{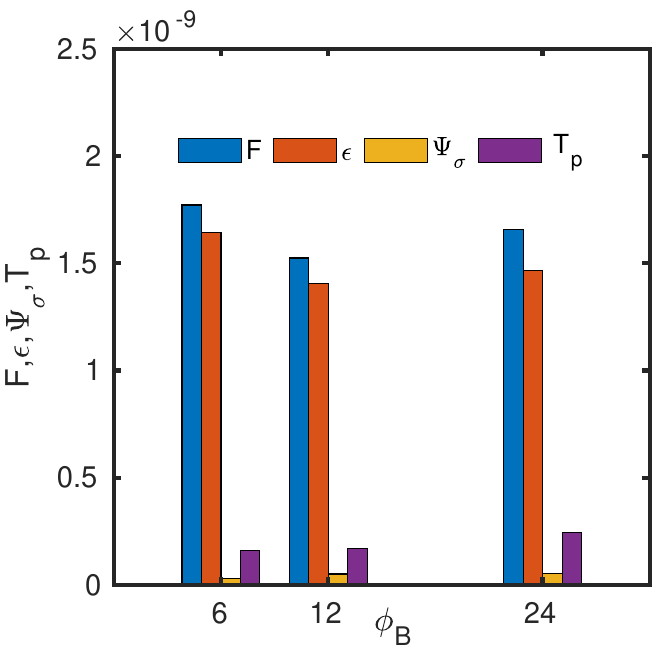}
    \caption{Global energy budget diagrams for increasing bubble volume fraction and for the baseline values of viscosity and density ratios. }
    \label{fig:en_budget}
\end{figure}

Useful information on the energy transfer mechanisms occurring in bubble-laden HIT, beyond spectral analysis, can be obtained by examining the global energy budget in statistically steady conditions for different bubble volume fractions \(\phi_B\) and for the baseline values of viscosity and density ratios, as reported in Figure~\ref{fig:en_budget}. 
The global energy budget equation reads as follows:
\begin{equation}
    \frac{dE}{dt}=F-\varepsilon+ T_p + \Psi_\phi ,
\end{equation}
where $E$ is the turbulent kinetic energy, 
$F=\langle \rho\,\mathbf{u} \cdot \mathbf{f} \rangle / \langle \rho \rangle$ is the power related to the external forcing used to sustain the turbulence in the flow, 
$\varepsilon=\langle 2 \rho \nu S_{ij} S_{ij}\rangle / \langle \rho \rangle$ is the rate of turbulent kinetic energy dissipation, 
$T_p=-\langle \mathbf{u}\cdot \nabla p \rangle / \langle \rho \rangle$ is the work of pressure forces per unit time, 
and $\Psi_\phi=\langle \mathbf{u}\cdot \mathbf{F}_\sigma \rangle / \langle \rho \rangle$ is the mean surface–tension power, already defined above. 
All terms in the present budget are computed using Favre averages, i.e.\ density–weighted quantities normalized by the domain-average density \cite{pandey2022turbulence,pandey2020liquid}

As evident in the figure, at the global level, the energy balance largely follows the classical single-phase turbulence picture, with the mean energy input by external forcing \(F\) balanced by the mean viscous dissipation \(\varepsilon\). 
The surface–tension power \(\Psi_\sigma\) is found to be very small compared to both forcing and dissipation (more than one order of magnitude smaller). 

As shown in \citep{dodd2016interaction, elghobashi2019direct}, the global average of $\Psi_\sigma$ in bubble- or droplet-laden turbulence is zero at steady state, when the surface area is statistically constant. However, in our configuration, \(\Psi_\sigma\) takes a slightly positive value, which can be attributed to the fact that we are using a diffuse interface approach for which the net dissipation is not zero \cite{Perlekar_2019}.

On the other hand, the pressure work, present when the density contrast is not one, computed here as the difference between $F$ and $\varepsilon$, represents a small fraction (roughly $10\%$) of the total energy. 
This is consistent with the fact that, from Gauss' theorem, the term $\int \mathbf{u}\cdot\nabla p \, dV = -\int p\,(\nabla\cdot\mathbf{u})\,dV$ vanishes in an incompressible fluid, where $\nabla\cdot\mathbf{u}=0$ (the velocity field is solenoidal within the bulk). Thus, we expect this term to be non-vanishing only near gas-liquid interfaces in the presence of strong density  contrasts.

The above observations suggest that in bubble-laden turbulence, energy is redistributed across scales and phases via interfacial stresses, with the dispersed gas phase acting as a site of localized energy injection and the liquid phase serving as the primary reservoir of dissipation. 
Surface–tension contributions, while playing a role in local intermittency and spectral energy redistribution, do not lead to significant net energy production or loss when integrated over the domain, although they are responsible for phase-specific modifications in the energy transfer process.

\subsubsection{Third order structure functions in bubble-laden HIT}
To gain a deeper understanding on  the mechanisms shaping small-scale turbulence, it is instructive to analyze the third-order turbulent invariant \cite{mccomb1995theory}:
\begin{equation}
S_3(r) = \langle (\delta u_r)^3 \rangle,
\end{equation}
where $\delta u_r = u(x+r) - u(x)$ is, the longitudinal velocity increment at distance $r$.
In classical Kolmogorov theory, $S_3(r)$ follows a linear scaling in the inertial range (i.e. $S_3(r)=-\frac{4}{5} \varepsilon r$ being $\varepsilon$ the energy dissipation rate), which implies that the energy transfer rate is scale invariant within the inertial range or, in other words, that the turbulence exhibits a form of self-similarity across scales\cite{Pope_2000,mccomb1995theory}.  Deviations from the linearity may stand as a clear signature of a non-classical cascade, providing important insights into energy transfer mechanisms in bubble-laden flows.

\begin{figure}
\centering
\includegraphics[width=0.46\linewidth]{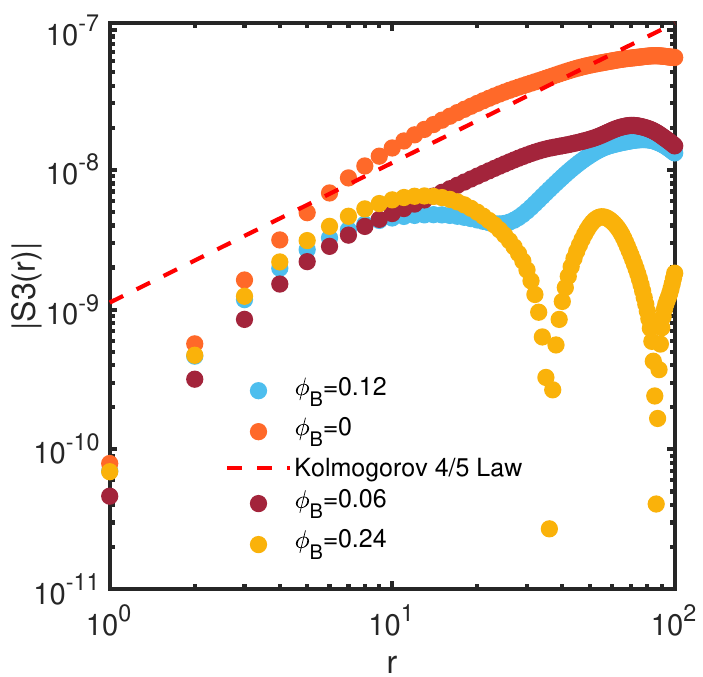}
\caption{Third-order structure functions, $S_3(r)$, for the cases under investigation. The data pertaining the single-phase flow are shown as a reference, along with Kolmogorov’s theoretical $4/5$ scaling law.}
\label{fig:S3_230}
\end{figure}

As shown in Figure~\ref{fig:S3_230},  single-phase turbulence closely follows the Kolmogorov prediction for the larger separations $r$ (i.e., the small-wavenumber regime). However, in the presence of a dispersed gas phase, the shape and magnitude of $S_3(r)$ appear significantly altered, now with significant differences for the different values of $\phi_B$.

At $\phi_B = 6\%$, $S_3(r)$ remains largely consistent with a forward energy cascade,  although its magnitude is noticeably reduced compared to the single-phase flow, which points to the presence of extra dissipative mechanisms in the intermediate range of scales for bubble-laden flows.

As the bubble volume fraction increases ($\phi_B = 12\%$ and $24\%$), $S_3(r)$ further deviates  from the Kolmogorov scaling, dropping  below the single-phase curve, particularly at intermediate scales. This is an indication of the presence of a reduced net energy transfer across scale due to vortex stretching and nonlinear interactions as bubble concentration increases, which can be attributed to the disruption of the large-scale coherence by the presence of the bubbles.
As noted in previous studies, interfacial stresses provide the alternative path for energy transfer to the smallest scales.

\begin{figure}
    \centering    \includegraphics[width=0.46\linewidth]{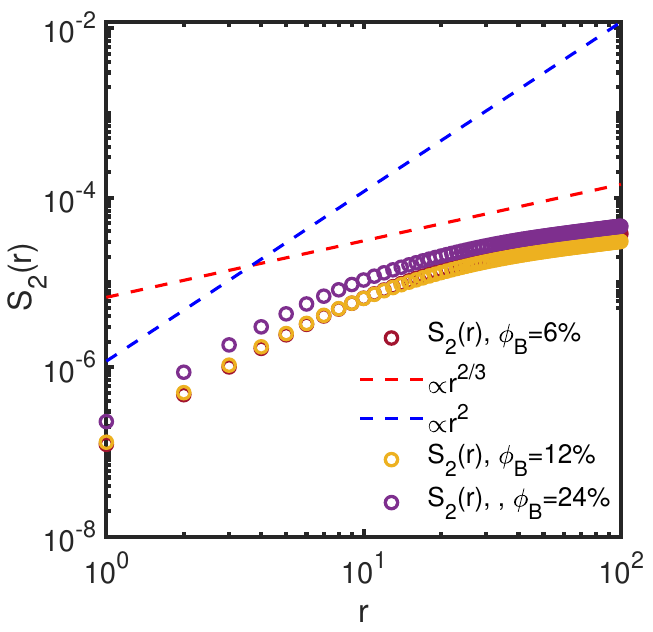}
    \caption{Second order structure functions for the baseline case and different bubble volume fraction. As a reference, the $\propto r^2$ and the $\propto r^{2/3}$ scaling laws are reported. }
    \label{fig:s2_multiphase}
\end{figure}

Note that, in figure \ref{fig:S3_230} we report the modulus of the third-order structure function and, consequently, $S_3(r)$ at $\phi_B = 0.24$ exhibits an oscillatory behavior with the presence of cusps at $r\sim 40$ and $\sim90$ (grid units).
The third-order structure function has been chosen in place of $S_2$ since the second-order function, being an even-order moment, is primarily sensitive to the energy distribution across scales, but not to the direction of the energy transfer. As such, it masks non-Kolmogorov dynamics such as non-local energy injection, backscatter, or anisotropy introduced by bubbles \cite{mccomb1995theory}. This can be readily observed in figure \ref{fig:s2_multiphase} reporting the $S_2(r)$ for the baseline case as a function of the bubble volume fraction.
The third-order structure function directly reflects the energy flux through scales and is therefore more sensitive to deviations from classical turbulence assumptions (e.g., scale locality, isotropy, and homogeneity). The observed departure in $S_3(r)$ therefore reveals the presence of altered cascade dynamics in bubble-laden flows that remain hidden in the second-order statistics.

The observations above show that the dispersed gas phase significantly modifies the turbulent energy transfer across scales, apparently suppressing a classical forward cascade at high volume fractions. In particular, the deviation from the theoretical $-4/5r$  behavior highlights the fundamental impact of gas-phase inclusions on the energy transfer in homogeneous and isotropic turbulence. However, the latter observation seems to contradict, at least partially, the results stemming from the inspection of the energy spectra, which still suggest the presence of a net forward cascade over a broad range of wave numbers. To reconcile this apparent discrepancy, we resort to the analysis of the energy transfer mechanisms within each phase separately. 

\subsubsection{Spectral energy distribution in continuous and dispersed phases: departure from Kolmogorov Scaling}

To understand the modifications of the energy transfer mechanisms in bubble-laden HIT, we examine the scale-by-scale energy dynamics within each phase. 

\begin{figure}
\centering
\includegraphics[width=0.7\linewidth]{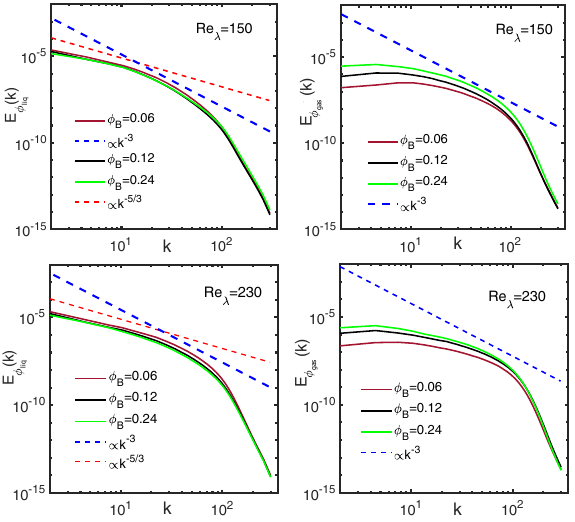}
\caption{Phase-specific energy spectra at $Re_\lambda=150$ and $230$. The left panels show the spectra of the liquid phase, while the right panels correspond to the dispersed gas phase.}
\label{fig:E_k_phasebyphase}
\end{figure}

As shown in Figure~\ref{fig:E_k_phasebyphase}, the energy spectrum in the liquid phase still follows the classical scaling at small-to-intermediate wavenumbers and transitions to a steeper regime (the dash reference line denote $k^{-3}$ scaling) before reaching the dissipation range. In contrast, the gas phase exhibits a nearly flat spectrum at low wavenumbers, followed by a $k^{-3}$ scaling and an even faster decay at smaller scales.
It should be noted that the spectra in the gas phase agree with those observed in \cite{cifani2020flow} and, interestingly, they closely overlap the analytical expression proposed by Risso \cite{risso2011theoretical} of a \( k^{-3} \) behavior in dispersed multiphase flows. In \cite{risso2011theoretical}, the author noted that the spectrum of a signal composed of a sum of localized random bursts can, under certain conditions, exhibit an intermediate sub-range that follows a power-law decay of \( k^{-3} \). For this behavior to emerge, the bursts should be characterized by a smooth and regular pattern, their strength and size should be statistically independent, and their size should be uniformly distributed between two finite wavelengths. 


The above results suggest that, despite the turbulence within each phase deviates from the Kolmogorov cascade, the global energy spectrum, as observed in the previous subsection,  still recovers the universal $k^{-5/3}$ across a wide range of wavenumbers.

Remarkably, the emergence of a $k^{-3}$ scaling in both the gas and liquid energy spectra occurs close to the Hinze scale, where the third-order structure function $S_3(r)$ also undergoes a sign inversion. This spatial and spectral coincidence indicates that the Hinze scale marks a transition point at which classical, scale-local energy transfer breaks down, and interfacial mechanisms, such as coalescence-driven recoil and bubble deformation, dominate the dynamics. These processes at the Hinze scale may be associated to the localized random bursts producing the  \( k^{-3} \) spectra \cite{risso2011theoretical}.
The sign change in $S_3(r)$ further suggests the presence of non-local or even reversed energy flux, highlighting the critical role of surface tension effects at this scale.

\begin{figure}
\centering
\includegraphics[width=0.45\linewidth]{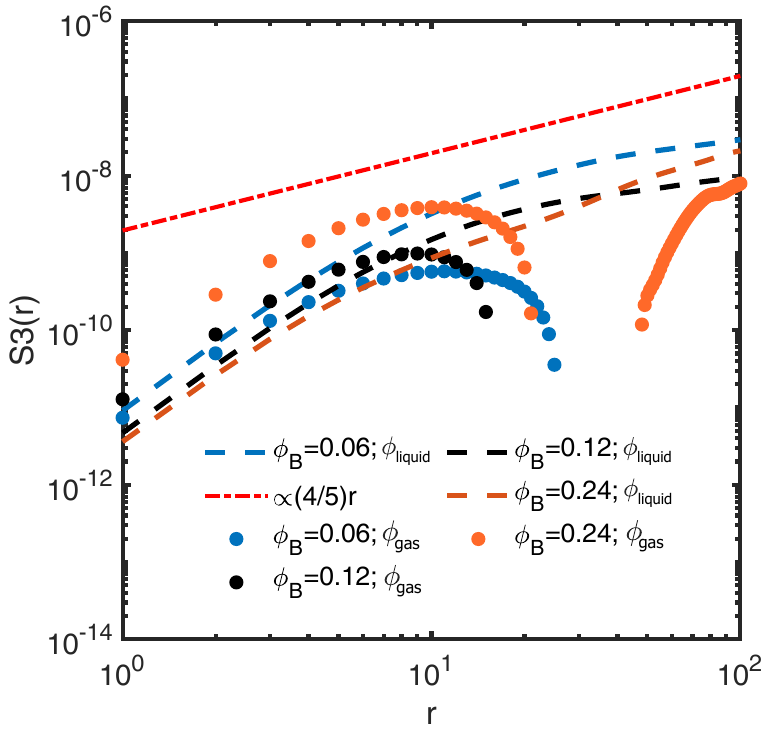}
\caption{Phase-specific third-order structure functions, $S_3(r)$, at $Re_\lambda=150$. The theoretical Kolmogorov $4/5$ scaling law is shown as a dashed red line. Circles represent $S_3$ in the liquid phase, while plusses correspond to the dispersed gas phase. $\phi_{liquid}$ and $\phi_{gas}$ correspond to the liquid and gas phases, respectively.}
\label{fig:s3_liquidgas_re150}
\end{figure}

Further insights into phase-specific energy transfer mechanisms can be gained by analyzing the phase-specific $S_3(r)$ structure functions. As shown in Figure~\ref{fig:s3_liquidgas_re150}, in the liquid phase, $S_3(r)$ largely follows the expected Kolmogorov scaling at large separations, confirming the fractal nature of turbulence in the continuous phase.
In contrast, the gas phase exhibits a distinctly different behavior. At small separations, $S_3(r)$ follows the expected dissipative-range scaling. However, at intermediate scales, it decreases abruptly, eventually becoming negative as $r$ increases. 

This anomalous behavior is to be attributed to the presence of the gas-liquid interface, which introduces an additional mechanism for energy redistribution. 


Furthermore, as the bubble volume fraction increases, the magnitude of $S_3(r)$ in the gas phase becomes progressively smaller, indicating a reduced net energy flux at intermediate scales. Such a suppression of energy transfer aligns with the spectral analysis, where the gas phase does not develop a well-defined inertial cascade, but rather follows a $k^{-3}$ scaling at intermediate wavenumbers. This confirms the more intuitive interpretation that turbulence in the dispersed phase remains constrained by interfacial effects, which limits the development of the classical turbulence features.

\begin{figure}
    \centering
    \includegraphics[width=0.7\linewidth]{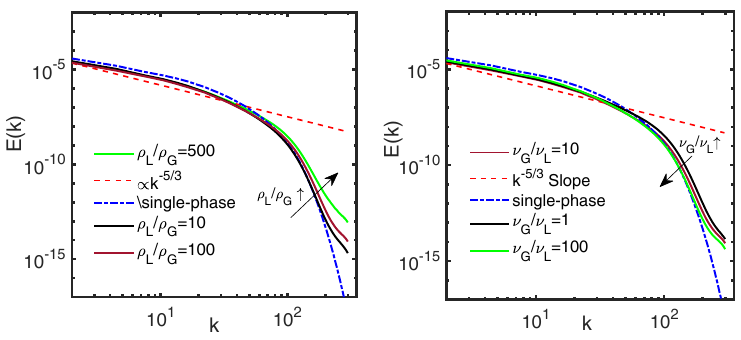}
    \caption{Energy spectra at $Re_\lambda=150$ at increasing density and viscosity ratio at $24\%$ gas volume fraction. }
    \label{fig:E_k_vardens}
\end{figure}

Taken together, the results from both the energy spectra and the third-order structure function analysis indicate that bubbles significantly modify turbulence by redistributing energy in a non-classical manner. While the liquid phase still supports a forward cascade, albeit weakened at intermediate scales, the gas phase experiences a fundamentally different energy transfer process, dominated by events at the intermediate scale, that we conjecture related to droplet breakup and coalescence. 

Summarising, the breakdown of Kolmogorov scaling may be attributed to a combination of factors intrinsic to the gas phase: confinement of turbulent eddies within finite-sized bubbles, low inertia, and surface tension-dominated dynamics. These effects likely suppress a scale-local energy cascade and instead promote non-local, intermittent transfer, associated with interface deformation, coalescence, or recoil. The presence of pronounced oscillations in $S_3(r)$, but not in the second-order structure function $S_2(r)$, further supports this view. While $S_2(r)$ captures the statistical energy content across scales, $S_3(r)$ encodes directional energy flux and reveals the underlying disruption or modification of the cascade mechanism specific to the gas phase.
All of the above highlights the critical role of inter-phase interactions in shaping turbulent energy dynamics in bubble-laden flows.

\subsubsection{Effect of density and viscosity ratio}

In this subsection we report on the separate effect of density and viscosity ratio on the modulation of the energy spectrum for $\phi_B=0.24$. The energy spectra, at $Re_\lambda=150$, are reported in Figure \ref{fig:E_k_vardens}. Both variations mainly affect the large and small wavenumber region of the spectrum, leaving almost unaltered the inertial cascade.

In particular, the increase of the density ratio, $\rho_L/\rho_G$ (the viscosity ratio is 1 in these cases), further reduces the slope of the energy spectra in the dissipative range,  i.e. smaller and smaller scales are excited in the flow when increasing the density ratio. The opposite effect is instead observed when increasing the viscosity ratio (the density ratio is fixed at $\rho_L/\rho_G=100$).  In any case, the observations made before still hold even in these cases, proving the robustness of the physical mechanisms described above. Indeed, by inspecting the third order structure function $S_3(r)$ for the case $\rho_L/\rho_G=500$, $\nu_G/\nu_L=1$ at $\phi_B=0.24$ (see fig.\ref{fig:S3_dens500} ), we observe that the Kolmogorov law is still clearly visible in the inertial range in the liquid phase while it is completely suppressed within the gas phase. Summarizing, the following conclusions may be drawn:

\begin{figure}
    \centering
    \includegraphics[width=0.4\linewidth]{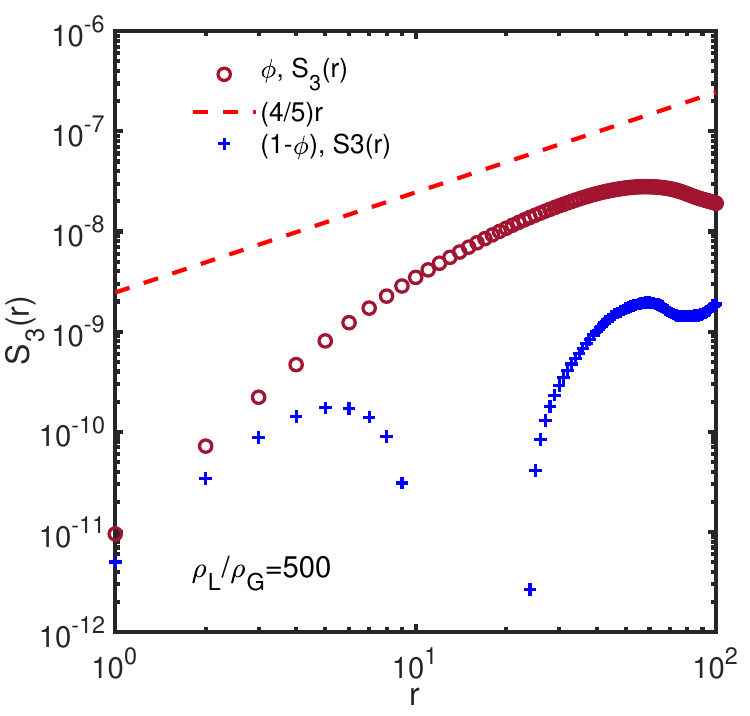}
    \caption{Phase-specific third-order structure functions, $S_3(r)$, at $Re_\lambda=150$, $\rho_L/\rho_G=500$ and $\phi_B=0.24$. The theoretical Kolmogorov $4/5$ scaling law is shown as a dashed red line. Circles represent $S_3$ in the liquid phase, while plusses correspond to the dispersed gas phase. $\phi$ and $1-\phi$ correspond to the liquid and gas phases, respectively.}
    \label{fig:S3_dens500}
\end{figure}

\textit{i) Universality holds in the Homogenized system}  

When considering the flow as a whole—i.e., analyzing the energy spectrum \(E(k)\) of the mixture without distinguishing between phases—the behavior closely follows the classical Kolmogorov prediction, at least for the configurations of this study. At large scales, within the inertial range, the presence of dispersed gas does not significantly alter the overall energy cascade in a way that is easily discernible from a homogenized perspective. 

\textit{ii)Phase-Specific Statistics Reveal Non-Kolmogorov Behavior}  

By decomposing the flow into liquid and gas phases and examining phase-specific energy spectra or third-order structure functions, deviations from classical turbulence emerge. In particular, the energy transfer within the gas phase diverges substantially from the  Kolmogorov scaling, with pronounced departures in small-scale statistics, precisely where multiphase effects become dominant (e.g., bubble interfaces, local slip velocities). In contrast, the liquid phase largely adheres to a near-classical scaling. This distinction implies that when recombining the two phases into a single energy spectrum, the non-Kolmogorov features primarily originating from the gas phase are partially masked.
%
Thus, while turbulence to the inertial-range scales remains close to classical phenomenology, the anomalous \(k^{-3}\) scaling emerges as a distinctive feature of the presence of the gas phase at smaller scales.

\subsubsection{Intermittency footprints in bubble-laden HIT}

\begin{figure}
    \centering
    \includegraphics[width=0.85\textwidth]{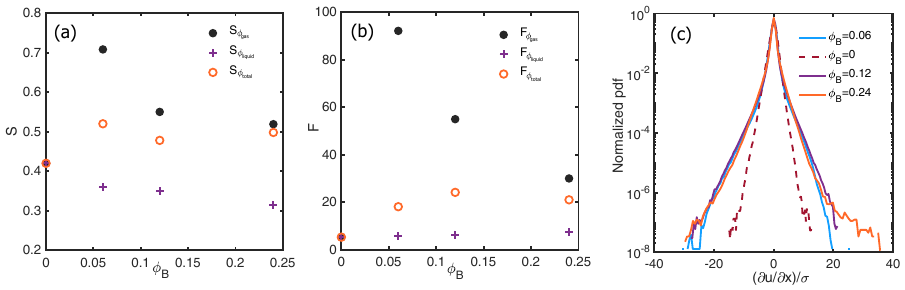}
    \caption{panel(a) Skewness \( S \) of velocity increments as a function of the bubble volume fraction \( \phi_B \). The values are reported for the liquid phase, the gas phase, and the homogenized system (subscript \textit{total}). panel(b) Flatness \( F \) of velocity increments as a function of the bubble volume fraction \( \phi_B \). The values are reported for the liquid phase, the gas phase, and the homogenized system. Panel (c) normalized PDF of the longitudinal velocity gradients at increasing bubble-volume fraction.}
    \label{fig:skewness}
\end{figure}

In the previous sections, we highlighted the presence of localized bursts of velocity within the gas phase, which point to strong intermittent events that alter the small-scale dynamics of bubble-laden turbulence. The statistical analysis of velocity derivatives provides further insights into this behavior. Figure~\ref{fig:skewness}(a) reports the skewness \( S \) of longitudinal velocity gradients as a function of the bubble volume fraction \( \phi_B \), while Figure~\ref{fig:skewness}(b) shows the corresponding flatness \( F \). 

In the liquid phase, both skewness and flatness remain close to the single-phase reference, confirming that the carrier fluid retains near-classical HIT statistics even in the presence of bubbles. In contrast, the gas phase shows pronounced deviations, especially at low void fractions. At \( \phi_B = 6\% \), the skewness reaches a peak of \( S \approx 0.7 \), while the flatness attains extreme values of \( F \approx 92 \), pointing to highly intermittent, heavy-tailed fluctuations. The probability density functions of the velocity gradients in Figure~\ref{fig:skewness}(c) further illustrate the enhanced occurrence of rare, intense events in the dispersed phase.  

Interestingly, these extreme values decrease with increasing gas volume fraction, suggesting that intermittency is most intense when bubbles are sparse and interfaces are more isolated. At higher \(\phi_B\), a more connected gas network appears to reduce the impact of rare events, although sampling limitations due to the reduced gas volume may also contribute to this trend.  

When the phases are combined into a homogenized system, the skewness stabilizes around \( S \approx 0.5 \) and flatness around \( F \approx 20 \). These aggregate values do not imply that the liquid mediates the gas-phase intermittency, but rather reflect the volume-weighted averaging of the two phases: the nearly Gaussian liquid statistics moderate the extreme gas-phase tails. As a result, the global system displays enhanced intermittency compared to single-phase HIT, but  weaker than the raw gas-phase signal.  

Overall, these results complement the spectral and structure-function analysis: the gas phase sustains strong non-Gaussian fluctuations at small scales, while the liquid phase remains close to Kolmogorov statistics. The intermittency diagnostics thus reinforce the view that gas–liquid interfaces act as sites of localized, intermittent energy transfers, but that their net contribution is partly obscured when examining homogenized turbulence statistics.

\section{Conclusions}

In this work, we investigated the dynamics of bubble-laden HIT through high-resolution simulations using a thread-safe, high-performance LB framework. The main focus is to shed light onto the modified energy transfer mechanisms and phase-specific deviations from Kolmogorov scaling in bubble-laden HIT.
We find that, at the level of the homogenized system, the energy spectra preserve a Kolmogorov-like behavior across a wide range of wavenumbers, confirming the apparent universality of turbulent energy transfer when phase distinctions are neglected. However, a phase-resolved analysis reveals that this global scaling masks important anomalies: the liquid phase largely maintains a classical turbulent cascade, while the gas phase exhibits non-Kolmogorov behavior, including a steeper $k^{-3}$ spectral decay and altered third-order structure functions.
These deviations stem from strong interfacial interactions and localized velocity bursts within the gas phase, indicative of pseudo-turbulent dynamics. The interface acts as both a source and sink of energy, disrupting the cascade and introducing intermittency, as evidenced by elevated flatness and skewness in the gas-phase velocity gradients. In contrast, the liquid phase shows only modest variations in intermittency, preserving the statistical features typical of HIT.

Our phase-specific energy flux analysis further highlights the fundamentally different roles played by the two phases. While the liquid supports a classical forward cascade, the gas phase lacks a clear inertial range and instead features localized, scale-selective energy injection and dissipation. Interestingly, the combination of these distinct behaviors still results in an aggregate energy spectrum that mimics Kolmogorov scaling.
Finally, we assessed the effect of varying density and viscosity ratios, confirming the robustness of our conclusions. Even under extreme contrasts, the qualitative features of energy redistribution and phase-specific cascade dynamics remain consistent.

In summary, our findings demonstrate that phase-specific interactions in bubble-laden turbulence not only disrupt classical scaling laws at small scales but also unveil fundamentally new turbulence dynamics that are obscured in global flow statistics. The distinct cascade dynamics in the gas and liquid phases --governed by localized interfacial bursts and non-classical energy transfers-- challenge the conventional universality of turbulent flows and call for a paradigm shift in the modeling of multiphase turbulence.  In future studies, we will extend our investigation to regimes involving buoyancy, near-contact interactions among fluid interfaces, and higher Reynolds numbers, aiming to bridge the gap between idealized turbulence and real-world multiphase phenomena.

\section*{Acknowledgements}

A.M. acknowledges funding from the Italian Government through the PRIN (Progetti di Rilevante Interesse Nazionale) Grant (MOBIOS) ID: 2022N4ZNH3 -CUP: F53C24001000006 and computational support of CINECA through the ISCRA B project DODECA (HP10BUBFIL). M.L. acknowledges the support of the Italian National Group for Mathematical Physics (GNFM-INdAM).


\bibliography{apssamp}
\end{document}